\documentclass[preprint,12pt,authoryear]{elsarticle}
\usepackage{epsfig}
\usepackage{amsmath}
\usepackage{graphicx}
\usepackage{cases}
\usepackage{color}
\usepackage{natbib}
\usepackage{psfrag}
\usepackage{amssymb}
\usepackage{tikz}
\usepackage{lipsum}
\usepackage{lineno}
\usepackage{cancel}

\def\be{\begin{equation}}
\def\ee{\end{equation}}
\def\bea{\begin{eqnarray}}
\def\eea{\end{eqnarray}}
\def\bml{\begin{mathletters}}
\def\eml{\end{mathletters}}

\journal{Journal of Theoretical Biology}

\begin{document}

\begin{frontmatter}



\title{Effect of drift, selection and recombination on the equilibrium frequency of deleterious mutations}
\author{Sona John \corref{cor1} and Kavita Jain\fnref{label2}}
\address{Theoretical
  Sciences Unit, \\Jawaharlal Nehru Centre for Advanced Scientific Research, \\ Jakkur P.O., Bangalore 560064, India\fnref{label3}}

\fntext[label2]{Corresponding author: \\ E-mail address: sonajohn@jncasr.ac.in (Sona John)}












\begin{abstract}
We study the stationary state of a population evolving 
under the action of random genetic drift, selection and recombination in which
both deleterious and reverse beneficial mutations can occur. 
We find that  the equilibrium
fraction of deleterious mutations decreases as the population size is increased. We calculate exactly the steady state frequency in a nonrecombining population when population size is 
infinite and for a neutral finite population, and obtain bounds on the fraction of deleterious mutations.  
We also find that for small and very large populations, the number of deleterious mutations depends weakly on recombination, 
but for moderately large populations, recombination alleviates the effect of deleterious mutations. An analytical argument shows 
that recombination decreases disadvantageous mutations appreciably when beneficial mutations are rare as is the case in adapting microbial 
populations, whereas it has a moderate effect on codon bias where the mutation rates between the  preferred and unpreferred codons are comparable. 
\end{abstract}

\end{frontmatter}


\section{Introduction}

A large number of population genetic studies assume one-way
mutation- in some situations, beneficial mutations are neglected as
they occur rarely \citep{Muller:1964,Felsenstein:1974,Haigh:1978,Gordo:2000a} while in adaptation studies,
deleterious mutations are 
ignored as they are unlikely to fix under strong selection
conditions \citep{Gerrish:1998,Rouzine:2008,Seetharaman:2014}. The assumption of one-way
mutation has an important effect on the nature of the state at 
large times. If the population size is infinite, a time-independent
stationary state can be reached due to a balance between mutation and
selection even if the mutational forces are unidirectional
\citep{Haigh:1978}. However in a finite population, when 
mutations are completely neglected or only 
unidirectional mutations are allowed, a  population evolving under the influence
of other evolutionary forces either does 
not reach an equilibrium state \citep{Haigh:1978}, or achieves a trivial one in which one
of the variants gets fixed at large times \citep{Ewens:2004}. It is when both beneficial and deleterious mutations are taken into account, a finite population reaches a nontrivial stationary state  \citep{Wright:1931}.  

An example of such a steady state is seen in the context of
synonymous codons that represent the same amino acid but do not occur
in equal frequencies \citep{Hershberg:2008,Plotkin:2011}. 
In a gene coding for a two-fold
degenerate amino acid, while   
selection favors the preferred codon, reversible mutations between
preferred and unpreferred codons and random 
genetic drift maintain the unpreferred  one \citep{Li:1987,Bulmer:1991}. 
Assuming that the sites in the sequence  
evolve independently, analytical results for the equilibrium frequency of 
unpreferred codons have been obtained
\citep{Li:1987,Bulmer:1991,Mcvean:1999}. 
However as the evolutionary dynamics at a genetic locus
are affected by other loci \citep{Hill:1966},   
a proper theory of codon usage 
bias must account for the Hill-Robertson interference between sequence
loci \citep{Comeron:1999,Mcvean:2000,Charlesworth:2009a,Kaiser:2009}. 

Reverse and compensatory mutations have also been proposed as
a possible mechanism to stop the degeneration of asexual populations 
\citep{Lande:1998,Whitlock:2000,Goyal:2012}. In a finite
nonrecombining population, if beneficial mutations are 
completely ignored,  
deleterious mutations accumulate irreversibly due to stochastic
fluctuations by a process known as Muller's ratchet
\citep{Muller:1964,Howe:2008}. But when rare beneficial mutations are taken into
account, the population reaches 
an equilibrium \citep{Estes:2003,Silander:2007,Howe:2008}.  
Recently \citet{Goyal:2012} calculated the amount of beneficial mutations required to achieve
a stationary state. But these authors assumed the mutation rates to be 
independent of the fitness, contrary to experimental evidence
\citep{Silander:2007}.  Moreover their solution for the equilibrium
frequency can become negative in some parameter range. 

In this article, we are interested in understanding the stationary
state of a multilocus model, which
is described in detail in the following section. We consider a class of non-epistatic fitness 
landscapes where the fitness depends only on the number of deleterious mutations
in a sequence ({\it{fitness class}}). 
As in previous works \citep{Li:1987,Comeron:1999,Mcvean:2000}, we assume that the beneficial mutations
are back mutations, the probability of whose occurrence depends on the
fitness class. More precisely, if the mutation probability per site is
small, the total probability of a beneficial (deleterious) mutation increases
(decreases) linearly with the fitness class. We consider the evolution of both infinitely large and finite populations, and to analyse the effect of
linkage amongst the loci, we allow recombination to occur. We are primarily 
interested in the population size dependence of the average number of disadvantageous mutations at
equilibrium. We obtain analytical results when the sites are   
completely linked, and compare them with the known results for a
freely recombining population. For intermediate recombination rates, we obtain numerical
results.

We find that the number of deleterious mutations decreases in a reverse
sigmoidal fashion, as the population size is increased. For small
populations, the fraction of disadvantageous mutations is seen to 
be roughly independent of population size and recombination rate. An
understanding of this behavior is obtained from an exact solution and numerical simulations  for a neutral finite population. For very large populations
that can be described by a deterministic model, we find the stationary
state exactly which is also unaffected by recombination. However for
moderately large populations, recombination is found to alleviate the effect of deleterious mutations \citep{Hill:1966,Felsenstein:1974,Barton:1998,
Charlesworth:2009a}, and the extent to which it does so depends on the beneficial mutation rate relative to the deleterious one. We find that when beneficial mutations 
are rare, the equilibrium frequency of disadvantageous mutations decreases logarithmically with population size 
when the loci are completely linked, but exponentially fast when linkage
is absent. On the other hand, when disadvantageous mutations are rare, the deleterious mutation fraction drops exponentially fast, irrespective of the recombination rate. 
Thus we expect that the linkage has a weak effect on codon bias where the rates at which mutations between preferred and unpreferred 
codons occur are of the  same order \citep{Zeng:2010,Schrider:2013}. 
But in adapting microbial populations where beneficial mutations are rare \citep{Sniegowski:2010}, recombination may be expected to reduce 
the frequency of disadvantageous mutations significantly.  


\section{Models}

We consider a haploid population of size $N$ in which each individual carries 
a biallelic (either zero or one) sequence of finite length $L$, 
where zero represents the wild type allele and one denotes the
deleterious mutation. The population is evolved in computer
simulations using a Wright-Fisher process in which recombination 
followed by mutation and selection occurs in discrete, non-overlapping
generations. To create an offspring, two parent individuals are
chosen at random with replacement. 
With probability $r \leq 1/2$, a single crossover event occurs in the parent
sequences at one of the $L-1$ equally likely break points to form 
two recombinant sequences, while with probability $1-r$, the parent
sequences are copied to the offspring sequences. 
In either case, one of the offspring  is chosen with
probability half to undergo mutations and selection, and the other one
is discarded. In the offspring sequence, a deleterious 
mutation occurs at a locus with a wild type
allele with probability $\mu$ and a reverse beneficial
mutation on mutant allele with probability $\nu$. 
The resulting sequence is allowed to survive with a
probability equal to its fitness, where the fitness of a sequence with
$j$ deleterious 
mutations is assumed to be a nonepistatic, and  given by $w(j)=(1-s)^j$, $0\leq s <1$.
This process is repeated until $N$ individuals in the next generation
are obtained. 

We have been able to implement the procedure described above for  
sequences of length up to $500$ and population sizes of the order
$10^3$. For larger populations with long nonrecombining sequence, the computational difficulties were
overcome by tracking 
only the number of deleterious mutations (fitness class) carried by 
the individual since the fitness of a sequence depends only on the
number of deleterious mutations in the 
sequence. Here a parent chosen at random produces a clone of itself, and the offspring may undergo mutations with a probability that 
depends 
on its fitness class. In a sequence with $j$ deleterious mutations,  
as a  deleterious (beneficial) mutation can happen at any one of the $L-j$ ($j$) sites, the
rate of  deleterious and  beneficial mutations is given by  $(L-j)\mu$ and $j \nu$ respectively. 
To find the number of beneficial ($b$) and deleterious ($d$) mutations
acquired by the offspring, random variables were drawn 
from Poisson distribution with mean $j \nu$ and $(L-j) \mu$
respectively. The total number of deleterious mutations in the offspring is then
given by $j'=j+d-b$.  If $j'$ turns out to be greater than $L$ or less
than zero, the offspring individual is produced with $j'=j$
mutations. 
As before, the offspring is allowed to survive with
  probability $w(j')$, and the process is repeated until  $N$
individuals in the next generation are obtained. 

All the numerical results presented here are obtained with an 
initial condition in which none of the individuals carry deleterious
mutations. In each stochastic run, the Wright-Fisher process was
implemented for about $10^4$ generations and it was ensured that the
stationary state is reached. In the  
equilibrium state of each run, we measured the number of deleterious mutations present in the population and 
averaged them over another $10^4$ generations. The data were also averaged over
$100$ independent stochastic runs.  
Although all the simulation results presented here are obtained using
the Wright-Fisher process, we will also use a continuous time Moran
model for some analytical calculations which is described in a later
section. 
If the population is infinitely large, the dynamics and equilibrium 
state of the population fraction can be described by a deterministic equation, which we discuss next.


\section{Results}

\subsection{INFINITE POPULATION}

\subsubsection{Nonrecombining population} For small selection coefficient and mutation rates, the population fraction ${X}(j,t)$ in the $j$th
fitness class at time $t$ evolves in continuous time according to 
\bea
 \frac{\partial {X}(j,t)}{\partial t} &=& -(s
 j+\overline{w}(t)){X}(j,t)-[(L-j) \mu + j \nu] {X}(j,t)\nonumber  \\
&+&(L-j+1) \mu {X}(j-1,t)   
 + (j+1) \nu {X}(j+1,t) ~,~0 \leq j \leq L
\label{contdtm}
\eea
where ${\overline w}(t) =  \sum_{k=0}^{L} \ln w(k) ~{X}(k,t) \approx -s \sum_{k=0}^{L} k ~{X}(k,t)$ is the
average Malthusian fitness and 
$X(-1,t)=X(L+1,t)=0$ at all times. In the above equation, the first term  
on the right 
hand side (RHS) represents the contribution to the change in
${X}(j,t)$ due to reproduction and the second term gives the loss in the
population fraction due to mutations. The last two
terms are the gain terms due to deleterious and beneficial mutations
respectively. 
The dynamics and the steady state solution of the deterministic model
defined by (\ref{contdtm}) can be found  
exactly. Below we discuss the 
stationary state and refer the reader to Appendix~A for the
time-dependent solution.

In the steady state, the left hand side (LHS) of (\ref{contdtm})
equals zero and  the population fraction carrying $j$ deleterious mutations is  of the
following product form \citep{Woodcock:1996}: 
\be
X(j)={L \choose j} ~x^j ~(1-x)^{L-j}
\label{prodx}
\ee
On using the above ansatz in
(\ref{contdtm}) for $j=0$ and $L$, we find that the  
average fitness ${\bar w}=L (\nu {\tilde x}- \mu)$ where ${\tilde x}=x/(1-x)$  
is a solution of the following quadratic equation:
\be
\nu {\tilde x}^2+ (s+\nu-\mu) {\tilde x}-\mu=0
\label{quad}
\ee
Plugging the ansatz (\ref{prodx}) in the bulk equations corresponding to
$j=1,...,L-1$ and rearranging the terms, we get 
\be
j \left( \mu-s-\nu + \mu {\tilde x}^{-1} -\nu {\tilde x}\right)-{\bar
  w}+ L (\nu {\tilde x}-\mu)=0 
\ee
which, by virtue of the results obtained above, shows that 
the ansatz (\ref{prodx}) is consistent with the bulk equations. Since
the population fraction must be positive, the allowed 
solution of (\ref{quad}) gives the fraction $x$ to be 
\bea
x &=& \frac{2 \mu}{\mu+\nu+s+\sqrt{(s+\nu-\mu)^2+4 \mu \nu}} 
\label{rform}
\eea
Furthermore, as the RHS of (\ref{prodx}) is a binomial distribution, 
the average fraction of deleterious mutations defined as $q={\bar
  j}/L=\sum_{j=0}^L j X(j)/L$ equals $x$. 

To get some insight in the solution obtained above, we first consider
some special cases by setting one of the parameters equal to zero. 

(i) In the absence of selection ($s=0$), we get 
\bea
X (j) &=& {L \choose j} \left(\frac{\mu}{\mu+\nu} \right)^j \left(\frac{\nu}{\mu+\nu} \right)^{L-j} \label{det_neu1}\\
q &=& \frac{\mu}{\mu+\nu}
\label{det_neu2}
\eea

(ii) When the reverse mutation probability $\nu$ equals zero, the
fraction $x=\mu/s~,~\mu < s$ and therefore 
\be
X(j)= {L\choose j} \left(\frac{\mu}{s} \right)^{j} 
\left(1-\frac{\mu}{s}\right)^{L-j} ~,~\mu < s
\label{nu0ss}
\ee
while for $\mu > s$, the fraction ${
  X}(j)=\delta_{j,L}$, thus signaling the well known error threshold
transition \citep{Wiehe:1997}. On the other hand, if the probability
$\mu$ is zero, we have the trivial solution that the fitness class with zero
deleterious mutations has frequency one, for all $\nu$. 

When all the three parameters are nonzero and the sequence length is large, the following cases may be considered \citep{Feller1:2000}:   

1. If $\mu, \nu, s$ are kept fixed but the sequence
length is increased, we find that the population 
fraction of deleterious mutations is a Gaussian centred about the average number $ L x$. 

2. If the deleterious mutation rate per genome $U_d=L \mu$ is held fixed while
$\mu \to 0, L \to \infty$, the fraction $x \approx \mu/(s+\nu)$ approaches
zero for finite $\nu$ and $s$. In this limit, the population fraction
is a Poisson distribution given by \citep{Pfaffelhuber:2012}
\be
{ X}(j) =
e^{-\frac{U_d}{s+\nu}}~\frac{1}{j!}~\left(\frac{U_d}{s+\nu} \right)^{j}
\label{poi}
\ee

3. However when both $\mu, \nu \to 0$ and $L
\to \infty$ such that the product $U_d=L \mu , U_b=L \nu $ remains
finite, taking $\nu \to 0$ in (\ref{poi}), we
immediately find that the population fraction is {\it
independent} of the beneficial mutation rate. To 
understand this rather surprising result, we first note that when
beneficial mutations are completely absent, due to (\ref{nu0ss}), the mean number
of deleterious mutations is of order unity {\it i.e.} it does not increase with $L$. However when beneficial mutations are present,  
the average number of advantageous mutations that can occur is  $\sim {\bar j}
\nu$  which approaches zero  as $\nu \to 0$, and thus the population remains unaffected by
beneficial mutations.

\subsubsection{Recombining population} So far, we discussed the stationary state of the deterministic model
when recombination is absent. But in an
infinitely large population, if epistasis is absent (as is the case 
here), the linkage disequilibrium (LD) stays at its initial value 
 \citep{Eshel:1970}. Since we start with an initially monomorphic
 population with zero LD, the results obtained above are expected to hold in a
 recombining population as well. In fact, when the sequence loci are
 completely unlinked ($r=1/2$) \citep{Nowak:2014}, \citet{Bulmer:1991} 
 has shown that the average fraction of deleterious mutations is given by (\ref{rform}).


\subsection{FINITE POPULATION WITHOUT SELECTION}

\subsubsection{Nonrecombining population} We consider a neutral Moran
process for an asexual population of finite size with a 
mutation scheme which is more general than that described in MODELS. 
In this model, a parent is randomly chosen with replacement to
replicate. If the offspring has $j$ mutations relative to the wildtype, the number of mutations 
increases (decreases) by one with probability $\mu_j$ ($\nu_j$) and remains unchanged with probability $1-\mu_j-\nu_j$. It is obvious that
$\mu_L=\nu_0=0$. An individual in the parent population is then
randomly chosen to die and is replaced by the possibly mutated
offspring. As explained in the Appendix~B,  
the average number ${\bar n}(j)$ of individuals carrying $j$ mutations evolves  according to (\ref{avg_Moran}). 
In the stationary state, we obtain 
\be
\frac{{{\bar n}(j)}}{N}= \frac{1}{1+ \sum_{k=1}^L \prod_{i=0}^{k-1} \frac{\mu_i}{\nu_{i+1}}} ~ \prod_{i=0}^{j-1} \frac{\mu_i}{\nu_{i+1}}
\label{app_avgfrac}
\ee
which is {\it independent} of the population size. 

\begin{figure}[h!]
\centering
\includegraphics[width=1\textwidth,angle=0]{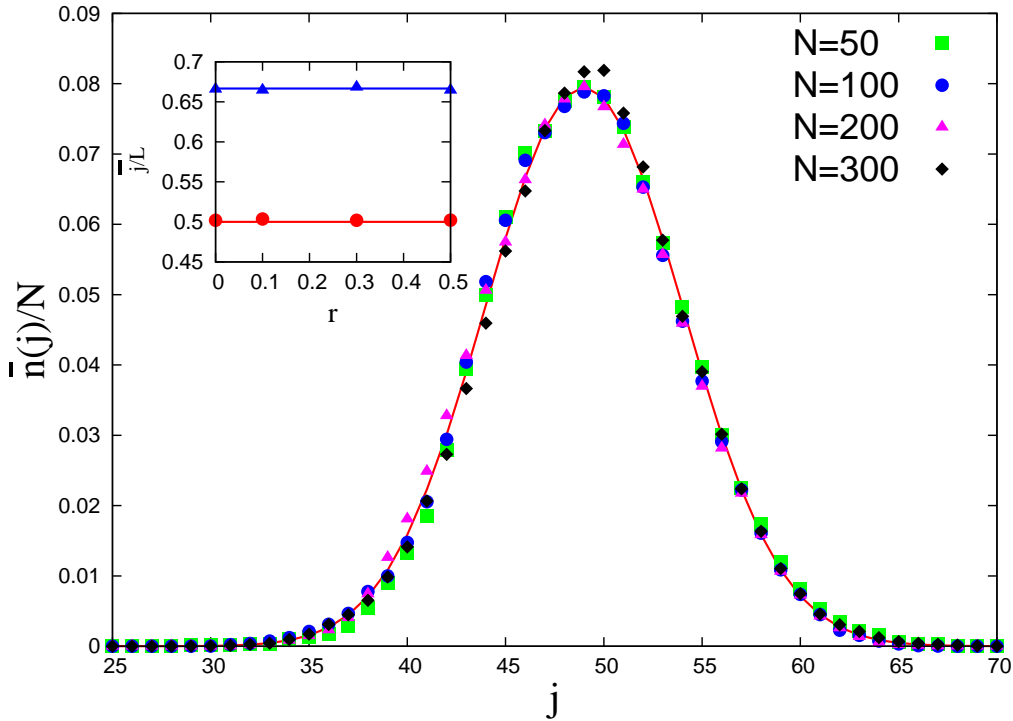}
\caption{Neutral case: Main figure shows the steady state fraction
  ${\bar n}(j)/N$ in the mutant class $j$ with $\mu=4.9
  \times 10^{-5}$, $\nu=5.1 \times 10^{-5}$ and $L=100$. The
  distribution is independent of the population size $N$, and matches 
  with the deterministic solution (\ref{det_neu1}) shown by solid line.    
The inset shows the average fraction of mutations relative to the wildtype
as a function of recombination probability $r$ for $L=300$, $N=300$
and $\mu=10^{-4}$ when $\nu=\mu$ ($\bullet$) and $0.5 \mu$
($\blacktriangle$). The solid lines give the 
theoretical prediction (\ref{det_neu2}).}
\label{fig_neu}
\end{figure}

For the model with back mutations, as explained in MODELS, 
the probability $\mu_j=(L-j) \mu$ and $\nu_j=j \nu$. Using this in the
above equation, we find that the average population fraction carrying
$j$ mutations is given by the deterministic solution
(\ref{det_neu1}) and the average fraction $q={\bar j}/L$  by
(\ref{det_neu2}), where ${\bar j}=N^{-1} \sum_{j=0}^{L} j {\bar n(j)}$. These results are verified in numerical simulations of the
Wright-Fisher process and are shown in Fig.~\ref{fig_neu}. 

\subsubsection{Recombining population} When the recombination probability is
equal to half, as the sequence loci evolve independently, the 
results from single locus theory are expected to hold. 
In this case, the frequency of  mutations is given exactly
by \citep{Wright:1931,Durrett:2008}   
\be
{\bar j}_1=\frac{\mu}{\mu+\nu}
\label{Durrett}
\ee
Thus the average number of mutations 
in the two limiting cases, namely for a 
nonrecombining population ($r=0$) and a freely recombining one
($r=1/2$), is same. Furthermore, the results
of our numerical simulations displayed in the inset of Fig.~\ref{fig_neu} for $0 \leq 
r \leq 1/2$ show that the average fraction $q$ is independent
of the recombination probability. 


\subsection{FINITE POPULATION UNDER SELECTION}

\subsubsection{Effect of sequence length} Our numerical simulations show that, unlike in the deterministic case, the fraction of deleterious mutations initially varies with the sequence length and approaches a constant value for long enough sequences. Motivated by the discussion for the
deterministic model, we consider the three cases when the sequence length is large. 

1. The limit in which $\mu, \nu, s$ are kept fixed but the sequence length is increased has
been studied in previous works to gauge the effect of Hill-Robertson
interference on the fraction of deleterious mutations
\citep{Comeron:1999,Kaiser:2009} and to understand the effect of
nonrecombining regions of different lengths in the genome of various
species \citep{Comeron:1999,Campos:2012}. 
Here for a given $N s$, the average fraction of deleterious mutations is found to
increase with increasing sequence length, but saturates to a
finite constant smaller than unity for long sequences. Our simulation data (not shown) 
is also consistent with this observation. 

2. When $U_d$ and
$\nu$ are kept finite and sequence length is increased, our simulations show that for long enough sequences,
 the average number of deleterious mutations ${\bar j}$  is a constant, as in the deterministic model.

3. In the rest of the article, we will consider the biologically relevant 
limit in which the genome mutation rates $U_b$ and $U_d$ remain
finite, as the number of loci in the sequence is increased
\citep{Drake:1998}. We find that unlike in the deterministic case, here the average {\it fraction} of 
deleterious mutations is finite and sensitive to the beneficial mutation rate. Figure \ref{kbarL} 
shows that the fraction $q$ decreases to a constant value, as the sequence length is increased. 
The data shown in the other figures of this article refers to this large-$L$ limit.

\subsubsection{Nonrecombining population} The neutral Moran model 
described in the last section can be straightforwardly generalised to 
include selection, but we find  that the evolution equation for the
average number distribution ${\bar n}(j)$ does not close in the presence of
selection {\it i.e.} it involves quantities that can not be 
expressed in terms of ${\bar n}(j)$. Therefore to understand the population size dependence of the average frequency
 $q$ of deleterious mutations, 
we use the results 
obtained in the last two sections, and employ an analytical argument which is described below.

\begin{figure}[t]
\includegraphics[width=0.9\textwidth,angle=0]{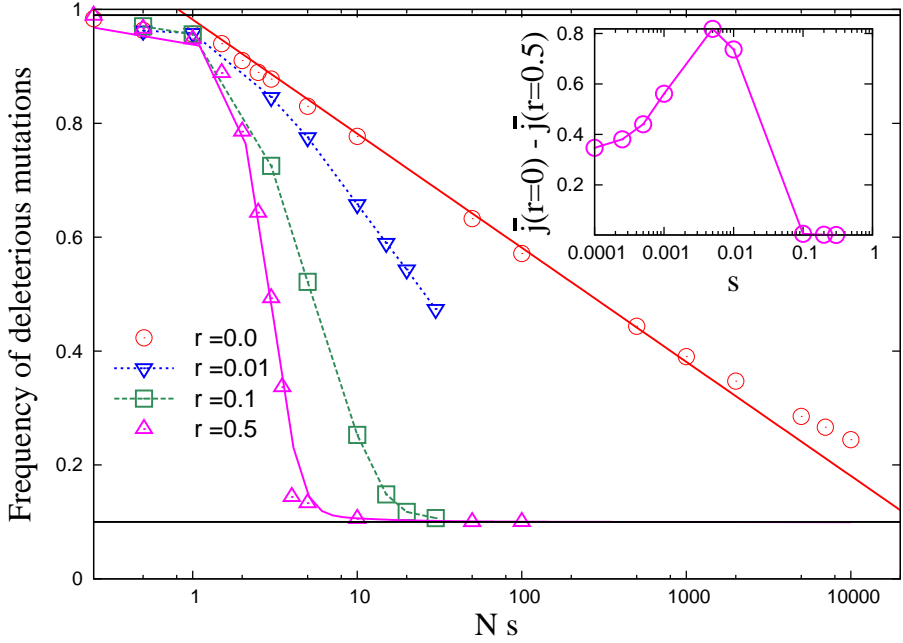}
\caption{Directional selection and rare beneficial mutations:
Figure shows the average frequency of disadvantageous mutations as a
  function of population size when $U_d=10^{-1}$, 
$ U_b= 10^{-3} $, $s= 10^{-2}$ and $L=100$. For nonrecombining population, the line shows the best fit curve  $ 0.087 \ln(N s)+ 0.98$ 
to the numerical data and for freely recombining population, (\ref{KiCrMa}) is shown. The broken lines joining the numerical 
data for $r=10^{-2}$ and $r=10^{-1}$ are a guide to the eye. 
The solid line at the bottom is the deterministic expression (\ref{rform}) and the one at the 
top shows the prediction (\ref{det_neu2}) from the neutral theory. The inset shows the nonmonotonic behavior of the difference 
between the deleterious mutations in a nonrecombining and freely recombining population for $N=1000, L=100, U_d=10^{-1}$ and
$U_b=10^{-3}$.}
\label{fig_Ubsmall}
\end{figure}

\begin{figure}[t]
\centering
\includegraphics[width=0.9\textwidth,angle=0]{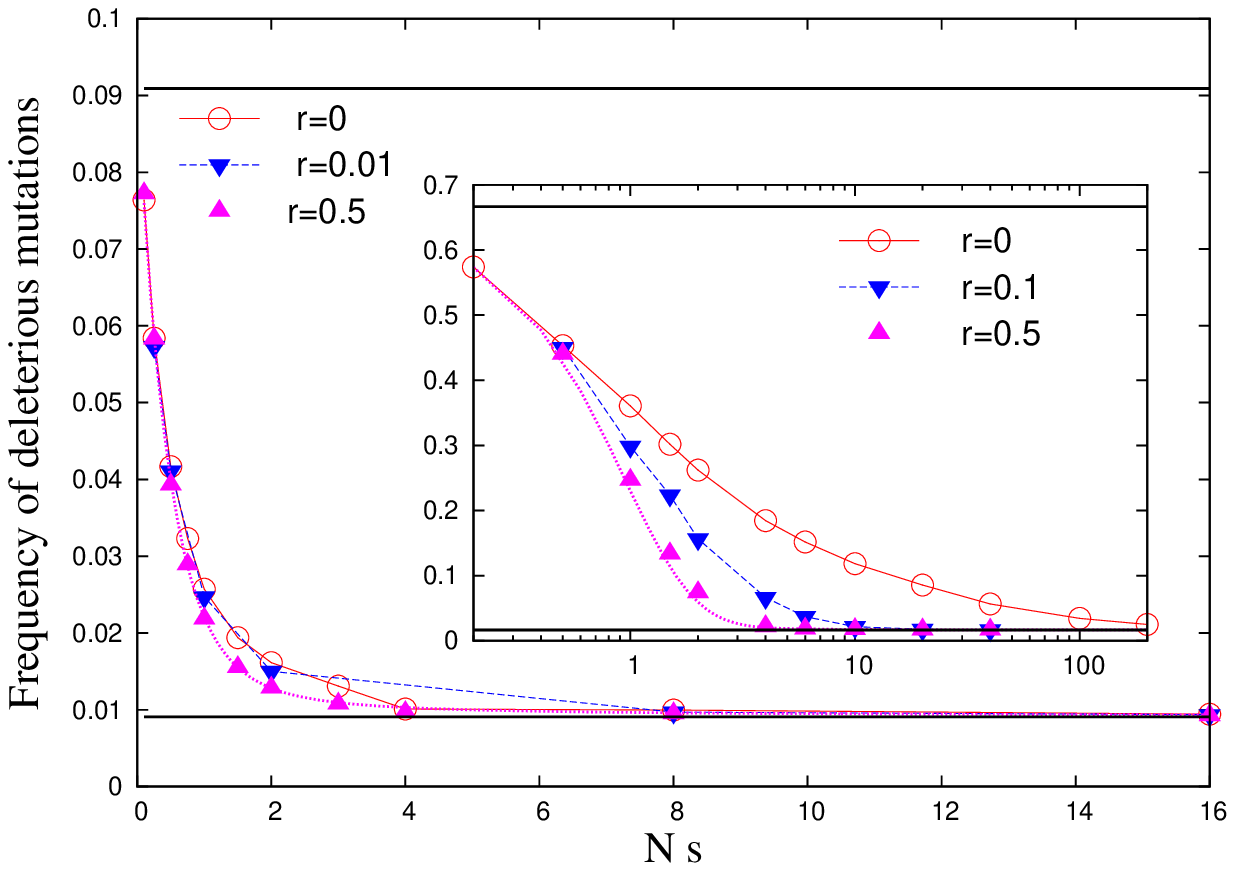}
\caption{Directional selection and frequent beneficial mutations: Figure shows the average frequency of disadvantageous mutations as a
  function of population size when $L=100, s=10^{-2}, U_d=10^{-2}, U_b=10 U_d$ (main) and 
  $L=300, s=2\times10^{-2},U_d=10^{-1}, U_b=0.5 U_d$ (inset). For freely recombining population, 
(\ref{KiCrMa}) is shown while rest of the curves are a guide to the eye.  
  The solid line at the bottom is the deterministic expression (\ref{rform}) and the one 
at the top shows the prediction (\ref{det_neu2}) from the neutral theory.}
\label{fig_Ublarge}
\end{figure}

{\underline {Small and very large populations:}} Figure~\ref{fig_Ubsmall} and~\ref{fig_Ublarge} show that
the fraction of disadvantageous mutations 
decreases monotonically with the population size $N$. When the
selection is weak ($N s \ll 1$), the fraction $q$ is expected to be close to the neutral value
(\ref{det_neu2}), in agreement with the data in
Figs.~\ref{fig_Ubsmall} and ~\ref{fig_Ublarge}. For very large
populations, the deterministic solution (\ref{rform}) is expected to
hold, and Fig.~\ref{fig_Ublarge} clearly shows that this
expectation is borne out by numerical simulations. 

\begin{figure}[h!]
\centering
\includegraphics[width=0.6\textwidth,angle=270]{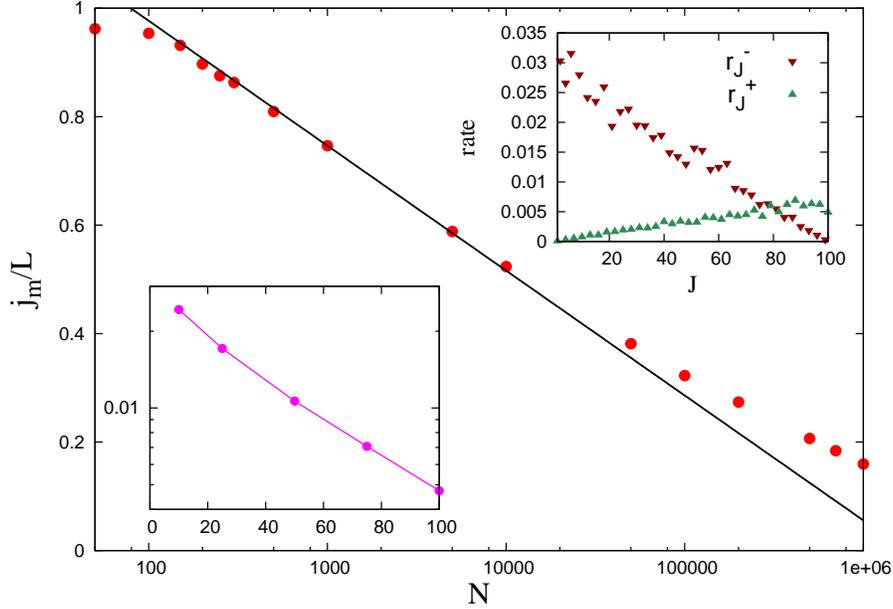}
\caption{Behavior of least-loaded class: Main figure shows the logarithmic decrease
 of the deleterious mutation fraction $j_m/L$
with population size $N$ when beneficial mutations are rare. The theoretical prediction (\ref{lnn})
 is shown with a best fit for the intercept as $1.437$. The parameters are $s=10^{-2}, U_d=10^{-1},U_b=10^{-3}$ and $L=100$.
Bottom, left inset: Plot to show the exponential decay of $j_m/L$ as predicted from (\ref{expN}) when
beneficial mutations occur frequently. The line joining the points is a guide to the eye. Here $U_b=10^{-1}$ and $U_d=10^{-2}$,
and the other parameters are same as those in the main figure.
Top, right inset: Degeneration and regeneration 
rates calculated in numerical simulations starting from all the
individuals in the best and worst fitness class respectively.
Parameters: $L=100, N=50, U_d=5\times 10^{-2}, U_b=5\times 10^{-3}$ and  $s=10^{-2}$.}
\label{Ndepn}
\end{figure}

{\underline {Moderately large populations:}} We now discuss a rate matching argument that allows us to find the {\it minimum} number
$j_m$ of deleterious mutations in the population.  
The basic idea is that if beneficial mutations are neglected,  
due to stochastic fluctuations, all the individuals in the least-loaded fitness class $j_m$
will acquire deleterious mutations and it will get lost from the population at a degeneration rate
$r_{j_m}^{-}$ \citep{Muller:1964,Haigh:1978}. However due to beneficial back
mutations, this process can be reversed and the population in the
fitness class $j_m$ will be regenerated at a rate $r_{j_m}^{+}$. In
the stationary state, on equating these two rates, the least-loaded fitness
class $j_m$ can be found \citep{Goyal:2012}. The variation of these
rates with the fitness class is shown in the inset of Fig.~\ref{Ndepn}, and we
observe that with increasing number of deleterious mutations, the degeneration rate decreases while the regeneration rate increases.  
This is a direct consequence of the fact
that for the fitness-dependent mutation scheme considered here (refer
MODELS), the total deleterious mutation rate $(L-j) \mu$ decreases with increasing $j$, but the beneficial mutation rate $j \nu$ decreases with decreasing $j$.

In the absence of beneficial mutations, as shown in 
Appendix~C,  the average number of individuals 
in the least-loaded fitness class $J$ is given by $n_J=N X_J^{(0)}(J)=
N (1-\mu/s)^{L-J}$ 
which grows exponentially with $J$. 
As a result, an initially fast-clicking ratchet with $n_J s \ll 1$ crosses
over to a slow-clicking ratchet with $n_J s \gg 1$, when $n_J s$ is of order
unity  \citep{Haigh:1978,Jain:2008b}. Using a diffusion theory for the slow ratchet
\citep{Stephan:1993,Gordo:2000a,Jain:2008b}, we find that when $n_J
\gg 1$, the degeneration rate is given by 
\be
r_J^-= \sqrt{\frac{N { X}^{(0)}_J (J) c^3 s^3}{\pi}}~ e^{-c s N { X}_J^{(0)} (J)}
\label{MRj}
\ee
where 
\be
{ X}_{J}^{(0)}(J) \approx e^{-\frac{U_d}{s} \left(1-\frac{J}{L} \right)}
\label{fracX}
\ee
and $c$ is a number of order unity \citep{Neher:2012,Metzger:2013}. 
When deleterious mutations are absent, a maladapted population adapts 
at a rate that depends on the number $N U_b$ of
beneficial mutants produced per generation. 
For $N U_b \ll 1$, the beneficial mutants arise one at a time and go to
fixation sequentially, while they interfere with each other for $N U_b
\gg 1$ \citep{Gerrish:1998}. 
The regeneration rate in these two parameter regimes is given by \citep{Park:2010,Goyal:2012}

\begin{numcases}
{r_J^+ \sim}
2 s N U_b ~(J/L)  ~,~N U_b \ll 1   \nonumber \\  
\frac{s \ln N}{\ln^2 U_b} ~ \frac{f(J)}{L}~,~N U_b \gg 1
\label{succ}
\end{numcases}
where, our numerical simulations for large populations indicate that $f(J)$ is of the form
 $\delta_1 \sqrt{J}+\delta_2 J$. The above equation shows that the rate $r_J^+$ depends weakly on $N$,
 and increases linearly with $J$ for large $J$.  

i. Rare beneficial mutations ($U_b \ll U_d$): An expression for $j_m$ can be obtained by matching the
rates  (\ref{MRj}) and (\ref{succ}). 
But as the degeneration rate decays fast with $N$ whereas regeneration
rate depends weakly on population size, we may treat the rate
$r^+_{j_m}$ as a constant 
in $N$. This simplification implies that $r_{j_m}^- \sim e^{-c s N { X}_{j_m}^{(0)} (j_m)} \sim
1$ which immediately leads to   

\be
\frac{j_{m}}{L} \sim -\frac{s}{U_d} \ln (Ns)
\label{lnn}
\ee
Our analytical result (\ref{lnn}) is compared with the results
of numerical simulations in Fig.~\ref{Ndepn} and for a wide range of
population sizes, we see a good agreement.  Figure~\ref{Xbark} shows that the average population fraction is 
distributed over a narrow range of
fitness classes \citep{Li:1987}, and therefore we may expect ${\bar j}$ to behave in a manner similar to $j_m$. 
Indeed as shown in Fig.~\ref{fig_Ubsmall},
 the average fraction of disadvantageous mutations also decreases logarithmically with population size,
 albeit with a prefactor smaller than $s/U_d$. 

ii. Frequent beneficial mutations ($U_b \gg U_d$): When $U_b \gg U_d$, 
the average frequency of deleterious mutations lies between the neutral value $\mu/\nu$ (refer (\ref{det_neu2})) and the deterministic value 
$\mu/(s+\nu)$ (refer (\ref{rform})), and thus $q \ll 1$ for a wide range of population sizes.
 This implies that $j_m/L$ is also small compared to unity. Using this in (\ref{fracX}), and that the degeneration rate $r_{j_m}^+$ 
is linear in $j_m$, we have 
\be
\frac{j_m}{L} \sim  e^{-c s N e^{-U_d/s}}
\label{expN}
\ee  
which decreases exponentially fast with population size and is consistent with our numerical observations shown in the inset of Fig. \ref{Ndepn}.
Same behaviour is observed for the average fraction of deleterious mutations ${\bar j}/L$, refer  Fig.~\ref{fig_Ublarge}.

\subsubsection{Recombining population} Having discussed the case of complete linkage ($r=0$), we now turn to the
limit of completely unlinked loci ($r=1/2$) where single locus theory
applies. When selection is present, a diffusion theory calculation 
\citep{Wright:1931} gives the frequency of deleterious mutations for a haploid population 
to be \citep{Kimura:1963} 
\be
{\bar
  j}_1=\frac{\mu}{\mu+\nu}~\frac{_1F_1(2 N \mu+1,2 N (\mu+\nu)+1
  ,-2 N s)}{_1F_1(2 N \mu,2 N (\mu+\nu),-2 N s)} 
\label{KiCrMa}
\ee
where $_1F_1(a,b,z)$ is the confluent hypergeometric function. For
$s=0$, the above expression reduces to (\ref{Durrett}). When $N s$ is
small, we have 
\be
{\bar j}_1=\left(1+\frac{\nu}{\mu} e^{2 N s}\right)^{-1}
\label{LiBu}
\ee
 which may be obtained either from (\ref{KiCrMa}) \citep{Li:1987,Bulmer:1991,Kondrashov:1995,Lande:1998} or a rate matching
argument \citep{Bulmer:1991,Lande:1998}. When $N s$ is large,
(\ref{KiCrMa}) approaches $\mu/s$ \citep{Kimura:1963} as one would
also expect from the deterministic solution (\ref{rform}). Thus as
Fig.~\ref{fig_Ubsmall} and \ref{fig_Ublarge} show, the fraction $q$ 
decreases exponentially fast in a reverse sigmoidal fashion, as the population size $N$ is increased when there is no linkage between loci.

As in the two extreme cases of complete linkage and no linkage, for $0 < r < 1/2$, 
we discern three distinct regimes in the behavior of the
fraction $q$ of disadvantageous mutations. Our numerical data in Figs.~\ref{fig_Ubsmall} and \ref{fig_Ublarge} 
shows that the fraction $q$ is roughly constant in population size and recombination rate when the 
population is small or very large. But for moderately large population, $q$ decreases with increasing population size and the
general effect of recombination is to decrease the equilibrium frequency of the deleterious mutations.

\section{Discussion}

In this article, we examined the stationary state of a model 
in which both  beneficial and
deleterious mutations can occur. The multilocus model studied here differs from that in 
previous works \citep{Gordo:2001,Goyal:2012} where these  mutation rates are assumed to be independent of the
fitness. Here we considered a biologically realistic situation of forward and backward mutations  where the rates depend linearly on 
the logarithmic fitness. In the general scenario where compensatory mutations can occur, nonlinear relationship
between the mutation rates and 
logarithmic fitness has been experimentally observed \citep{Silander:2007}. Here we are mainly concerned with the variation of the average number ${\bar j}$ of 
deleterious mutations with the population size. 

\subsection{Exact bounds on the number of deleterious mutations} For an infinitely large and nonrecombining population, exact results for the population frequency have been 
obtained for special choice of parameters  
\citep{Woodcock:1996,Maia:2003,Etheridge:2009,Pfaffelhuber:2012}, and here these results were generalised to obtain exact stationary \
state and dynamics. Since we consider  
non-epistatic fitnesses, the stationary state solution does not depend on the recombination rate \citep{Eshel:1970}. Moreover as the 
deterministic limit corresponds to very strong selection which is not favorable for disadvantageous mutations, this analysis provides a lower 
bound on the average number ${\bar j}$ of deleterious mutations. 

The upper bound on ${\bar j}$ can be found by considering the neutral limit for a finite population. For completely linked loci,
we calculated the average frequency of  mutations (relative to the wildtype) exactly, and found it to be independent of the population size. 
Although the latter result is known from previous studies on one locus models \citep{Durrett:2008}, 
to our knowledge,
 such a result has not been obtained using 
a multilocus model. Using numerical simulations and the known results for freely 
recombining population \citep{Wright:1931,Durrett:2008}, we found that the number ${\bar j}$ is independent of the recombination rate in the 
neutral limit as well. This happens because in the absence of selection, as random genetic drift creates positive and negative linkage disequilibrium (LD) with equal
 probability, the average LD vanishes \citep{Hill:1968,Hadany:2008} and therefore the average number ${\bar j}$ is not affected by recombination. 
It should however be noted that the higher moments of the number of  mutations may depend on 
both the recombination rate and population size \citep{Hill:1968}.

\subsection{Effect of drift, selection and recombination} To get an insight into the problem when both 
selection and population size are finite and recombination is absent,  
we used a rate matching argument which states that stationarity is achieved when the rate at 
which the least-loaded fitness class is lost due to deleterious mutations 
equals the rate at which it is regenerated by beneficial
mutations \citep{Goyal:2012}. A similar
argument has been used previously by \citet{Bulmer:1991}, but  
in a single locus setting, to arrive at the equilibrium fraction of
deleterious mutations given in (\ref{LiBu}). In recent years, 
some analytical understanding of the rate at which an asexual 
population declines in fitness 
\citep{Stephan:1993,Gordo:2000a,Jain:2008b,Etheridge:2009,Waxman:2010,Neher:2012,Metzger:2013} and adapts
\citep{Gerrish:1998,Wilke:2004,Rouzine:2008,Desai:2007a,Park:2010} has become
available in multilocus models. Using these results and the rate
balancing argument described above, we found analytical 
expressions for the minimum number of deleterious mutations that a
finite asexual population under selection carries in the stationary
state.  
 
For a nonrecombining population, our main result is that the average fraction $q$ of deleterious mutations decreases from the neutral value
(\ref{det_neu2}) 
towards the deterministic fraction (\ref{rform}), as population size is
increased. If beneficial mutations are rare ($U_b\ll U_d$), as is the case in
adapting microbial populations \citep{Sniegowski:2010}, $q$ changes
logarithmically with population size. In an adaptation experiment on
bacteriophage, it was observed that when the population size is
increased by a factor ten, the logarithmic fitness increased mildly
\citep{Silander:2007}, which is consistent with the weak
$N$-dependence seen here. 
Experimental data \citep{Zeng:2010,Schrider:2013} on Drosophila shows that the mutation rate 
from preferred to unpreferred codon is twice as much as that for the reverse mutations. In such a case where $U_b\sim U_d$, as 
the inset of Fig. \ref{fig_Ublarge} indicates, $q$ decreases faster than the logarithm of population size, but
we do not have an analytical form for it. However in the extreme case when $U_b \gg U_d$, we find that the fraction $q$
decreases exponentially fast with the population size. Similar qualitative behaviour, namely,  
the decrease in ${\bar j}$
with increasing population size is seen when recombination is nonzero, refer Figs. \ref{fig_Ubsmall} and \ref{fig_Ublarge}. 
When the population size is kept fixed and the selection 
coefficient is increased, the average fraction of deleterious mutations decreases as one would intuitively 
expect (data not shown). Although the rate balancing argument used here explains the population size dependence of 
the fraction of deleterious mutations, we have not been able to obtain a complete analytical understanding of its 
variation with selection coefficient since the $s$-dependence of the function $c$ in the degeneration rate in (\ref{MRj}) 
is not known. We also performed numerical simulations keeping the product $N s$ constant ($= 10$), and find that ${\bar j}$ 
is not a function of $N s$ unlike the one locus theory prediction (\ref{KiCrMa}). For $s = 0.005$, we obtained ${\bar j}= 13.7$
which increased to $28.8$ on halving $s$ which suggests that it depends more strongly on $s$ than $N$ which is consistent
with (\ref{lnn}).

For a given $N s$, we find that the recombination reduces the
frequency of the deleterious mutations (also see \citep{Barton:2010}). As discussed above, in a finite population, due
to random genetic drift, both positive and negative LD are created. If LD is positive, the population 
consists of individuals with extreme fitnesses on which selection can 
act efficiently and thus removes the LD. On the other hand, when LD is negative, as most of the individuals are likely to 
have similar fitnesses, selection is ineffective in removing LD. Thus in the presence of selection, the average LD in a 
nonrecombining population is negative 
\citep{Felsenstein:1974,Hadany:2008}. But once recombination is introduced,
it will create individuals with extreme fitnesses thereby helping selection to weed out the deleterious mutations, 
and thus decreasing ${\bar j}$. The effect is large for intermediate values of $N s$ 
since this regime corresponds to both selection and drift having a strong effect. From the results in the neutral 
and deterministic limit, we expect that the difference in the number of deleterious mutations carried by a nonrecombining
and recombining population is nearly zero when $s \ll 1/N$ (weak selection) and  $s \gg 1/N$ (strong selection). 
Thus, as shown in the inset of Fig.~\ref{fig_Ubsmall}, the maximum advantage of recombination occurs at an intermediate 
value of selection coefficient as has also been observed in other studies \citep{Gordo:2008}.

Although recombination reduces the number of deleterious mutations, the extent to which it does so depends on how 
common the beneficial mutations are compared to the deleterious ones. In an adapting asexual population where beneficial 
mutations occur rarely \citep{Sniegowski:2010}, even slight 
recombination reduces ${\bar j}$ considerably indicating the advantage of recombination during adaptation 
\citep{Barton:1998,Hadany:2008}. On the other hand, in the codon bias problem where back mutation rates are comparable to the forward ones 
\citep{Zeng:2010,Schrider:2013}, the fraction of unpreferred codons is given by (\ref{LiBu}) if the loci are assumed to be completely unlinked, 
but as the inset of  Fig.~\ref{fig_Ublarge} shows, linkage increases the unpreferred codon frequency 
moderately \citep{Comeron:1999,Mcvean:2000,Charlesworth:2009a}.

\subsection{Effect of background selection - an application} Background selection is a type
 of Hill-Robertson 
 effect \citep{Charlesworth:2012} and is known to increase  
 the rate at which the Muller's ratchet clicks  \citep{Gordo:2001,Kaiser:2010}. 
  In a finite, nonrecombining population with an infinitely long 
 sequence in which both deleterious and beneficial mutations occur at $L$ {\it background selection sites},  
 and deleterious mutations accumulate at 
rest of the sites \citep{Kaiser:2010}, we find that the ratchet clicking time is considerably reduced from the 
situation when there are no background selection sites (see Fig.~\ref{Eff}). linIf the background selection 
sites (BGS) remain at equilibrium in the presence of other linked loci also, they affect the evolutionary dynamics at 
other sites, and their effect can be quantified by a reduction in the effective population size to the number of 
individuals carrying the minimum number of deleterious mutations at BGS \citep{Charlesworth:2012,Gordo:2001}. 
Since the minimum number of deleterious mutations in the BGS is $j_m$, we require the population fraction in the class 
$j_m$. For large populations with $N s \gg 1$ where the deterministic theory is expected to hold, using (\ref{poi}) 
and (\ref{fracX}), we obtain
\be
N_e= N e^{-\frac{U_d (1- \frac{j_m}{L})}{s}}
\label{eff}
\ee
where $j_m$ is a function of population size $N$. The ratchet
time with background selection for a population of size $N$ is found to be well
approximated by the ratchet time without it for a population of size
$N_e$ as shown in Fig.~\ref{Eff}. From the results for $j_m$ when $U_b \ll U_d$, we 
expect $N_e$ in (\ref{eff}) to increase linearly with 
$N$ for small and large populations. But for the intermediate range of
population sizes, using (\ref{lnn}) in (\ref{eff}) above, we find the 
effective population size to be independent of $N$. These predictions were  
tested numerically and as shown in Fig.~\ref{Eff}, the effective population size and the ratchet time remain roughly 
constant when the 
actual population size is varied over three orders of magnitude . 

\section{Conclusions}

We close this article by listing some open questions. Here we investigated the effect of linkage using numerical simulations, 
but an analytical expression for the average ${\bar j}$ 
as a function of recombination probability is desirable. We also considered the specific case of forward and backward mutations, and an 
extension of these results to the more general case of compensatory mutations would be interesting.

{\bf{Acknowledgement}} 

The authors thank B. Charlesworth, M. M. Desai,
S. Goyal, J. Krug and L. M. Wahl for discussions, and are grateful to
B. Charlesworth for useful comments on an earlier version of the manuscript.



\newpage

\bigskip
{\bf{\centerline{Appendix A: Deterministic dynamics and stationary state}}}
\label{app_detdyn}
\bigskip

Equation (\ref{contdtm}) is nonlinear in the population fraction due
to the first term on the RHS. This nonlinearity can 
be eliminated by a change of variables from $X(j,t)$ to an unnormalised 
population variable $Z(j,t)$ which is defined as \citep{Jain:2007b,Jain:2011c}
\be
Z(j,t)= X(j,t)~e^{\int_0^t dt' ~{\overline w}(t')} 
\ee
Then the unnormalised population fraction obeys the following {\it
  linear} differential equation:
\bea
\frac{\partial {Z}(j)}{\partial t} &=& -s j~
     {Z}(j,t) -[(L-j) \mu + j \nu] {Z}(j,t) \nonumber  \\  
 &+& (L-j+1) \mu {Z}(j-1,t) + (j+1) \nu {Z}(j+1,t) 
\label{Zcontdtm}
\eea
with boundary conditions 
\be
{Z}(-1,t) = {Z}(L+1,t) =0
\label{bc}
\ee
at all times. The RHS of (\ref{Zcontdtm}) is a three-term recursion
relation (in $j$) with variable 
coefficients, which is usually not easy to solve. 

Inspired by the results of \citet{Woodcock:1996}, we assume that the
population fraction ${Z}(j,t)$ is of the following form
\bea
{Z}(j,t) 
= {L \choose j} ~x_1^{j}(t)~ x_2^{L-j}(t)
\label{ansatz}
\eea
where $x_1, x_2$ are calculated below. The normalised fraction 
$X(j,t)$ is then given by  \citep{Jain:2007b,Jain:2011c}
\bea
X(j,t) &=& \frac{Z(j,t)}{\sum_{j'=0}^L Z(j',t)} \\
&=& {L \choose j} x^{j}(t) ~(1-x(t))^{L-j} 
\label{normfrac}
\eea
where $x=x_1/(x_1+x_2)$ lies between zero and one. It should be noted 
that the above form for the population fraction of a fitness class implies that each locus in the sequence contributes
{\it independently} to the population fraction of a sequence.

Using the ansatz (\ref{ansatz}) in the
boundary conditions (\ref{bc}), we 
find that $x_1, x_2$ obey linear, coupled differential
equations which can be expressed as 
\be
\frac{\partial}{\partial t}\begin{pmatrix} {x}_1 \\ {x}_2  \end{pmatrix}=
\begin{pmatrix} -\nu-s  & \mu \\ \nu & -\mu \end{pmatrix} ~\begin{pmatrix} x_1 \\ x_2  \end{pmatrix}
\label{r1r2}
\ee
On using the ansatz (\ref{ansatz}) in the bulk equation
(\ref{Zcontdtm}) for which $0 < j < L$, we get   
\be
\frac{j}{x_1} \left[\frac{{\partial x}_1}{\partial t}+\left(\nu+s  \right)
x_1- \mu x_2 \right]+\frac{L-j}{x_2} \left[\frac{{\partial x}_2}{\partial t}  -\nu x_1
  +\mu  x_2 \right]=0 
\label{matrix}
\ee
However due to (\ref{r1r2}), the coefficient of $j$ and $L-j$ equals
zero for any $0 <j < L$. Thus the ansatz (\ref{ansatz}) is
consistent with the bulk equations, and the problem reduces to solving
the matrix equation (\ref{r1r2}). By going to the diagonal basis, 
we obtain 
\be
\begin{pmatrix} {x}_1(t) \\ {x}_2(t)  \end{pmatrix}=
\begin{pmatrix} \frac{2 \mu}{\nu-\mu+s+
  \sqrt{(\mu-s-\nu)^2+4 \mu \nu}}  & \frac{2 \mu}{\nu-\mu+s-
  \sqrt{(\mu-s-\nu)^2+4 \mu \nu}} \\ 1 & 1 \end{pmatrix} ~\begin{pmatrix} e^{\lambda_+ t} ~{\tilde x}_1(0) \\ e^{\lambda_- t} ~{\tilde x}_2(0)  \end{pmatrix}
  \label{r1r2b}
\ee
where the column vectors in the matrix above are the eigenvectors 
of the matrix on the RHS of (\ref{r1r2}) corresponding to the eigenvalues
$\lambda_\pm$, which are given by   
\be
\lambda_\pm = \frac{-\nu-\mu-s \pm
  \sqrt{(\mu-s-\nu)^2+4 \mu \nu}}{2}
\label{eigen}
\ee
and ${\tilde x}_1(0),{\tilde x}_2(0) $ can be found using the initial
condition $X(j,0)$. 

In the steady state, the population fraction  is obtained by taking
the limit $t \to \infty$ in the expressions of $x_1 (t), x_2 (t)$
obtained above. Using the fact that the eigenvalue $\lambda_-$ in
(\ref{eigen}) is negative, we find that the steady state 
fraction $x$ is given by (\ref{rform}). 

\bigskip
{\bf{\centerline{Appendix B: Moran model for neutral, nonrecombining population}}} 
\label{app_Moran}
\bigskip

For the Moran process defined in the main text, the probability distribution $P(n(i),t)$ of the number of individuals in the fitness class $i$ evolves according to the following equation:
\bea
&& \frac{\partial P(n(i),t)}{\partial t} \nonumber \\
&&=\sum_{j \neq i} \left[ \sum_{n(j)=1}^{N-n(i)}  P(n(i)+1,n(j)-1,t) ~R(n(i)+1 \to n(i),n(j)-1 \to n(j))  \right.\nonumber \\
&&- \sum_{n(j)=0}^{N-n(i)}  P(n(i),n(j),t) ~R(n(i) \to n(i)-1,n(j) \to n(j)+1)  \nonumber \\
&&+\sum_{n(j)=1}^{N-n(i)} P(n(i)-1,n(j)+1,t)  ~R(n(i)-1 \to n(i),n(j)+1 \to n(j)) \nonumber \\
&&- \left. \sum_{n(j)=0}^{N-n(i)}   P(n(i),n(j),t) ~R(n(i) \to n(i)+1,n(j) \to n(j)-1) \right]
\eea
where $R$ is the rate at which a birth-and-death event occurs, and $P(n(i),n(j),t)$ is the joint distribution of the number of individuals in the $i$th and $j$th fitness class.  Using the above equation, it can be seen that the average number of individuals in the fitness class $i$ given by ${{\bar n}(i,t)}=\sum_{n(i)=1}^N n(i) P(n(i),t)$ changes as 
\bea
\frac{\partial {{\bar n}(i,t)}}{\partial t}=\sum_{j \neq i} \sum_{n(i),n(j)} [{R}(n(i) \to n(i)+1,n(j) \to n(j)-1) P(n(i),n(j),t) \nonumber \\
 - {R}(n(i) \to n(i)-1,n(j) \to n(j)+1) P(n(i),n(j),t)]
\label{avg}
\eea

We next find the rates at which the birth-and-death process occurs. For class $i, j=0,...,L$ and $i \neq j$, we have 
\bea
&& R(n(i) \to n(i)+1,n(j) \to n(j)-1) \nonumber \\
&=&  (1-\mu_i-\nu_i)  \frac{n(i)}{N}  \frac{n(j)}{N} 
+ \mu_{i-1} \frac{n (i-1)}{N} \frac{n(j)}{N}+\nu_{i+1} \frac{n (i+1)}{N}  \frac{n(j)}{N} 
\eea
with $n(-1)=n(L+1)=0$. In the above equation, the first term on the RHS gives the probability of the event that a
 birth occurs in the $i$th class, the offspring does not mutate and a death occurs in the $j$th class,
 while the second and third term  give the probability that a birth occurs in a class neighboring the $i$th class,
 the offspring acquires a  mutation and a death occurs in the $j$th class.  On using the above equation
 in (\ref{avg}), after some simple algebra, we get
\be
\frac{\partial {{\bar n}(i,t)}}{\partial t}=\mu_{i-1} {{\bar n}(i-1,t)} + \nu_{i+1} {{\bar n}}(i+1,t)-
(\mu_i+\nu_i) {{\bar n}(i,t)} ~,~0 \leq i \leq L
\label{avg_Moran}
\ee
which can be easily solved in the stationary state to give (\ref{app_avgfrac}). 

\bigskip
{\bf{\centerline{Appendix C: Deterministic solution in the absence of beneficial mutations}}}
\label{app_diff}
\bigskip

Consider an infinitely large, nonrecombining population when only deleterious mutations are allowed. Let $J$ be the least-loaded fitness
class so that the frequency $X_J^{(0)}(k,t)=0~,~k < J$ at all times. 
Then the evolution equation (\ref{contdtm}) reduces to  
\bea
 \frac{\partial X_J^{(0)}(j,t)}{\partial t} &=& -(sj+{\bar w}_J (t)) X_J^{(0)}(j,t)-(L-j) \mu X_J^{(0)}(j,t) \nonumber \\
 &+& (L-j+1) \mu X_J^{(0)}(j-1,t)~,~J \leq j \leq L
\eea
where the average fitness ${\bar w}_J(t)=-s \sum_{k=J}^L k X_J^{(0)}(k,t)$. In the stationary state, the equation for $j=J$ gives
\be
{\bar w}_J=-(L-J) \mu -s J
\ee
On iterating the two-term recursion relation for $X_J^{(0)}(j)$, we obtain
\begin{eqnarray}
 X_J^{(0)}(j)
={L-J \choose j-J} \left(\frac{\mu}{s} \right)^{j-J} \left(1-\frac{\mu}{s} \right)^{L-j} ~,~\mu < s
\label{app3eqn}
\end{eqnarray}
For $J=0$,  (\ref{nu0ss}) is recovered. 


\newpage











\clearpage
\setcounter{figure}{0}

\begin{center}
{\bf{SUPPORTING INFORMATION}}
\end{center}

\begin{figure}[h!]
\centering
\renewcommand*{\thefigure}{S\arabic{figure}}
\includegraphics[width=0.8\textwidth,angle=0]{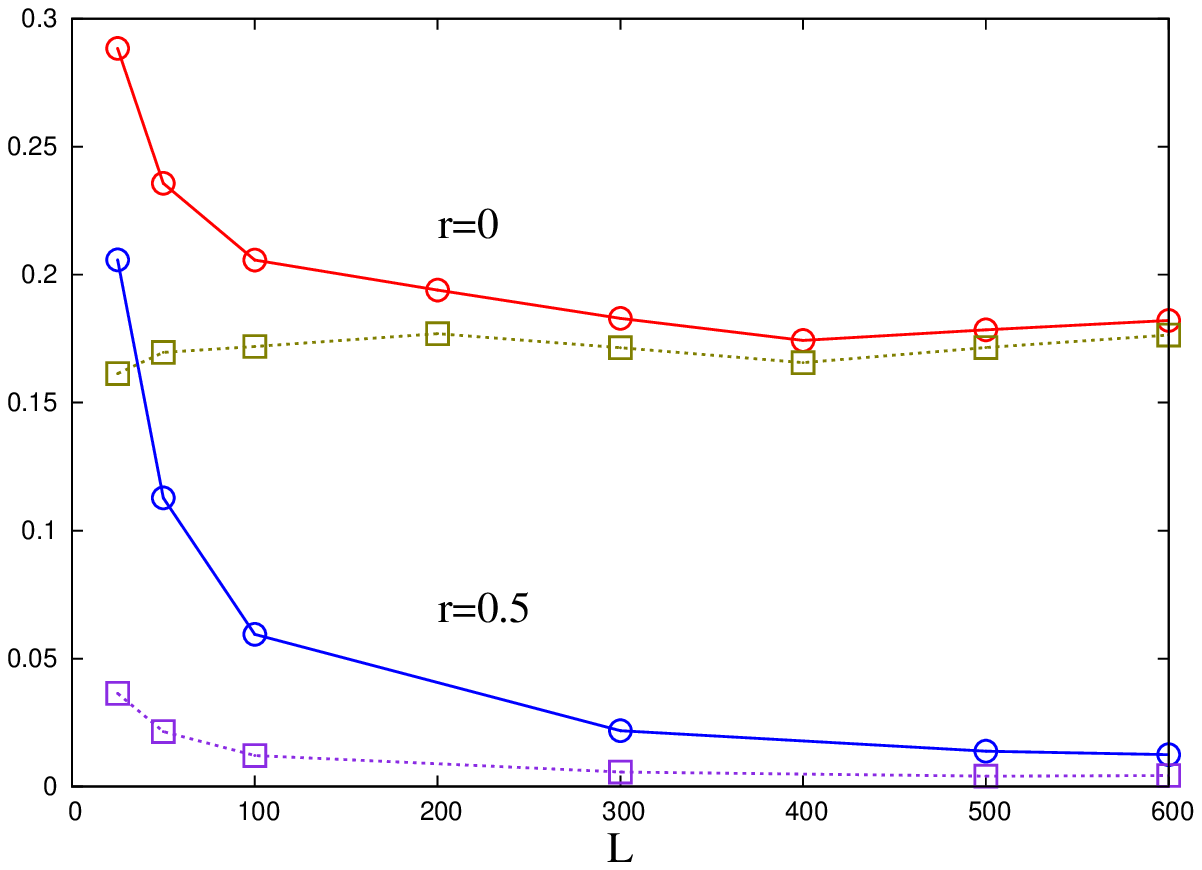}
\caption{{Variation of minimum number $j_m$ (broken line with $\square$) and average number  $\bar j$ (solid line with $\circ$)
 of deleterious mutations with sequence length $L$ for $N=200$, $s=2\times10^{-2}$, $U_d=10^{-1}$ and $U_b=5\times10^{-2}$.}}
\label{kbarL}
\end{figure}

\clearpage

\begin{figure}
\centering
\renewcommand*{\thefigure}{S\arabic{figure}}
\includegraphics[width=0.8\textwidth,angle=270]{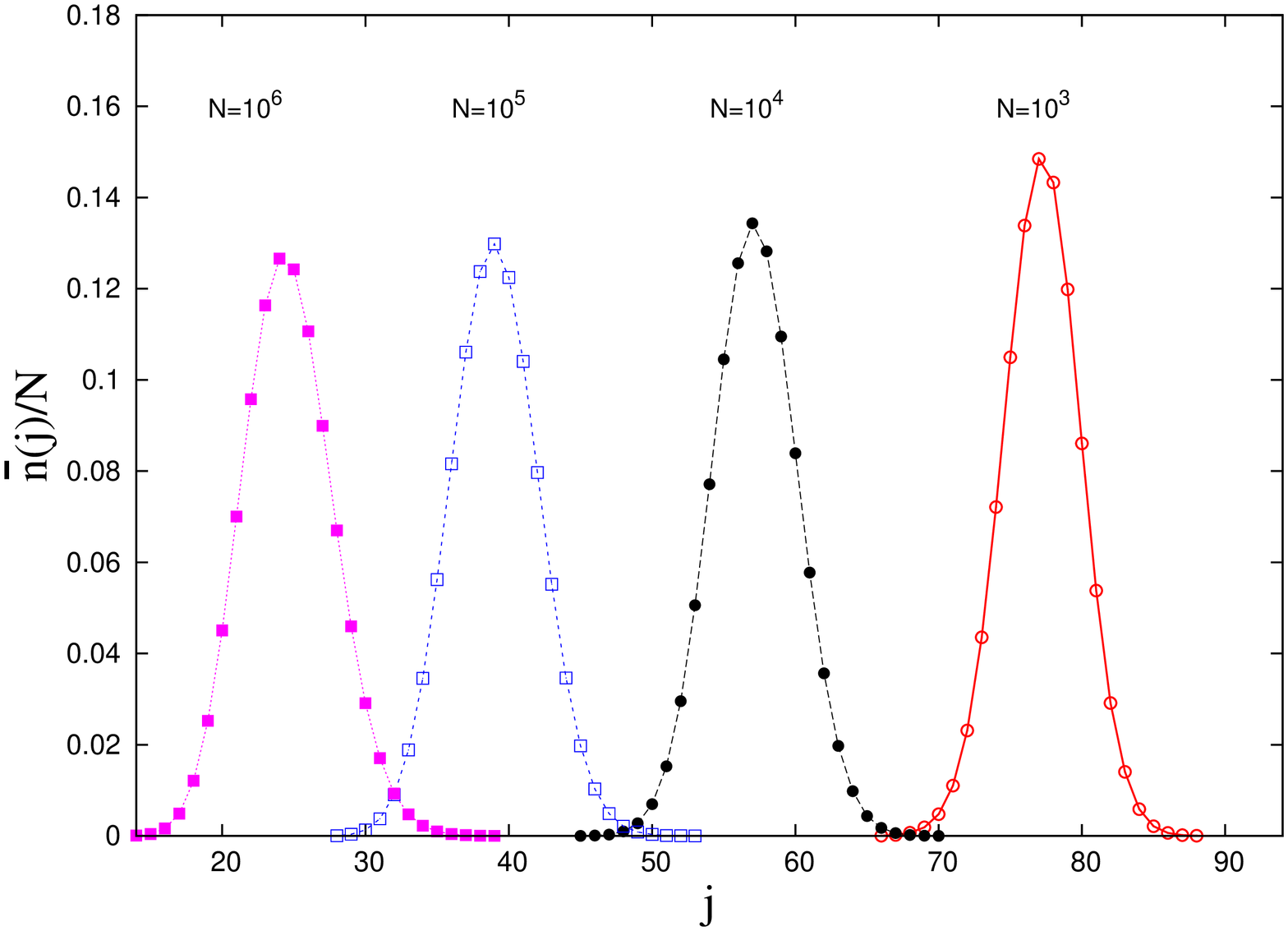}
\caption{Distribution of the average fraction of individuals in each mutant class for a completely linked sequence for various population sizes, and $L=100$, $U_d=10^{-1}$, $U_b=10^{-3}$ and $s=10^{-2}$.} 
\label{Xbark}
\end{figure}

\clearpage

\begin{figure}
\centering
\renewcommand*{\thefigure}{S\arabic{figure}}
\includegraphics[width=1\textwidth,angle=0]{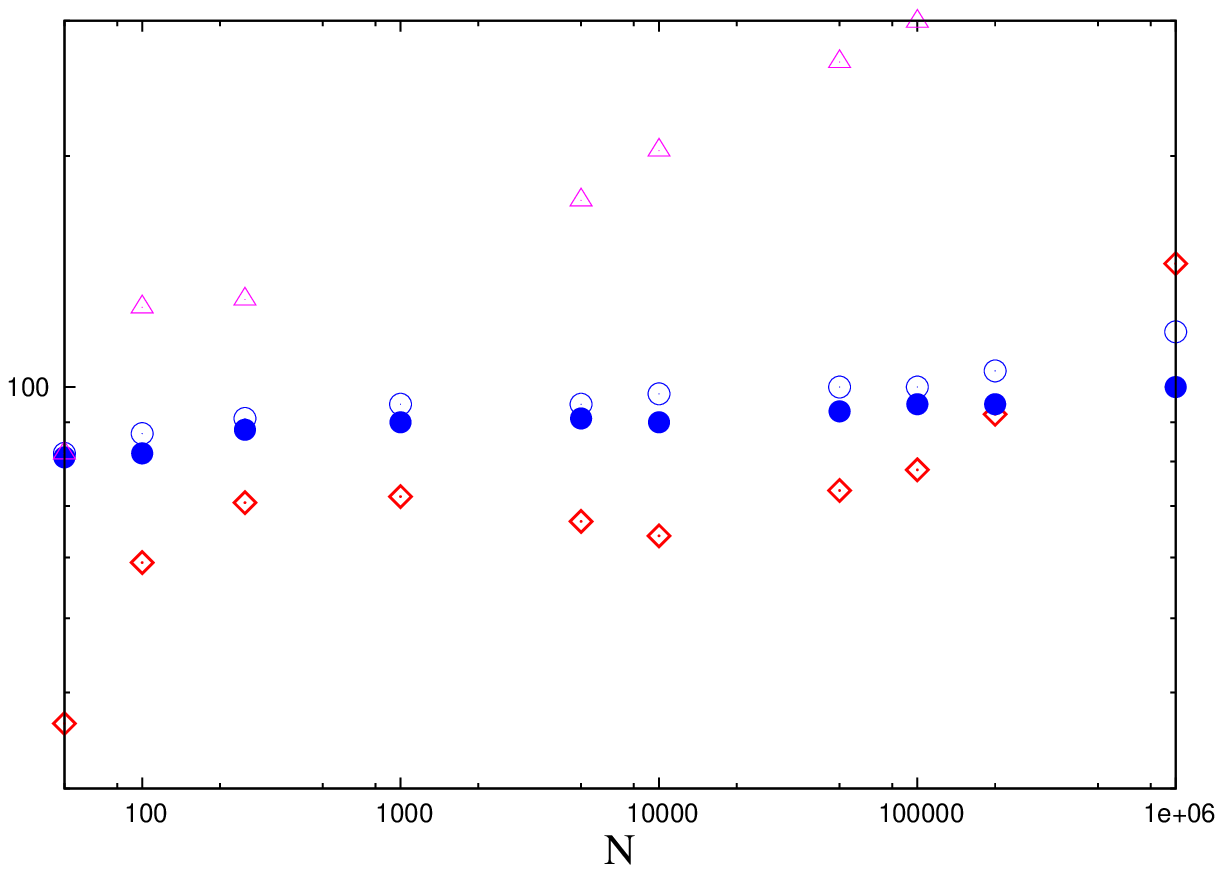}
\caption{Figure shows the effective population size $N_e$ ($\lozenge$) given by
  (\ref{eff}) where $j_m$ is shown in Fig.~\ref{Ndepn}, and click time of
  the Muller's ratchet with deleterious mutation rate $U_d'$ and selection coefficient $s'$, when there are finite number of background
  selection sites with reversible mutations. The click time of the
  ratchet for (i) a population of size $N$ and without background
  selection sites ($\vartriangle$), (ii)  a population of size $N$ with background
  selection sites ($\circ$) and (iii)  a population of size $N_e$ 
  without background selection sites ($ \bullet $) are shown. Parameters:
  $L=100,s=0.01, U_d=0.1, U_b=0.001,
  U_{d}'=0.015,s'=0.001$. }  
\label{Eff}
\end{figure}


\end{document}